\documentclass[12pt]{article}
\usepackage{graphicx}
\usepackage{color}
\usepackage{cite}

\def\upstrut{\vrule height 2.5ex depth 0.0ex width 0pt}

\def \beq{\begin{equation}}
\def \eeq{\end{equation}}
\def\eqref#1{(\ref{#1})}
\def\bea{\begin{eqnarray}}
\def\eea{\end{eqnarray}}
\def\jpsi{J\kern-0.15em/\kern-0.15em\psi\kern0.15em}

\def\URLtilde{\lower0.2em\hbox{$\tilde{\phantom{a}}$}}
\def\mycomm#1{\hfill\break\strut\kern-3em{\color{red}\tt ====> #1
\color{black}}\hfill\break}

\newcount\timecount
\newcount\hours \newcount\minutes  \newcount\temp \newcount\pmhours
\hours = \time
\divide\hours by 60
\temp = \hours
\multiply\temp by 60
\minutes = \time
\advance\minutes by -\temp
\def\hour{\the\hours}
\def\minute{\ifnum\minutes<10 0\the\minutes
\else\the\minutes\fi}
\def\clock{
\ifnum\hours=0 12:\minute\ AM
\else\ifnum\hours<12 \hour:\minute\ AM
\else\ifnum\hours=12 12:\minute\ PM
\else\ifnum\hours>12
\pmhours=\hours
\advance\pmhours by -12
\the\pmhours:\minute\ PM
\fi
\fi
\fi
\fi
}

\def\monthname{\relax\ifcase\month 0/\or January\or February\or
March\or April\or May\or June\or July\or August\or September\or
October\or November\or December\else\number\month/\fi}

\def\bold#1{\setbox0=\hbox{$#1$}     \kern-.025em\copy0\kern-\wd0
\kern.05em\copy0\kern-\wd0
\kern-.025em\raise.0433em\box0 }

\begin{document}
\setcounter{footnote}{1}
\rightline{EFI 16-22}
\rightline{TAUP 3012/16}
\rightline{arXiv:1611.00348}
\vskip1.5cm
\centerline{\large \bf \boldmath $QQ \bar Q \bar Q$ STATES:}
\medskip

\centerline{\large \bf MASSES, PRODUCTION, AND DECAYS \unboldmath}
\bigskip

\centerline{Marek Karliner$^a$\footnote{{\tt marek@proton.tau.ac.il}},
Shmuel Nussinov$^a$\footnote{{\tt nussinov@post.tau.ac.il}},
and Jonathan L. Rosner$^b$\footnote{{\tt rosner@hep.uchicago.edu}}}
\medskip

\centerline{$^a$ {\it School of Physics and Astronomy}}
\centerline{\it Raymond and Beverly Sackler Faculty of Exact Sciences}
\centerline{\it Tel Aviv University, Tel Aviv 69978, Israel}
\medskip

\centerline{$^b$ {\it Enrico Fermi Institute and Department of Physics}}
\centerline{\it University of Chicago, 5620 S. Ellis Avenue, Chicago, IL
60637, USA}
\bigskip
\strut

\begin{center}
ABSTRACT
\end{center}
\begin{quote}
The question of whether there exist bound states of two heavy quarks $Q=(c,b)$
and antiquarks $\bar Q = (\bar c, \bar b)$, distinct from a pair of
quark-antiquark mesons, has been debated for more than forty years. We estimate
masses of $Q_1 Q_2 \bar Q_3 \bar Q_4$ resonant states $X_{Q_1 Q_2 \bar Q_3
\bar Q_4}$
and suggest means of producing and observing them.  We concentrate on the
$c c \bar c \bar c$ channel which is most easily produced and the $b b \bar b
\bar b$ channel which has a better chance of being relatively narrow. We obtain
$M(X_{cc\bar c \bar c}) = 6{,}192 \pm 25$ MeV 
and
$M(X_{bb\bar b \bar b}) = 18{,}826 \pm 25$ MeV,
for the $J^{PC} = 0^{++}$ states involving charmed and bottom tetraquarks,
respectively.  Experimental search for these states in the predicted mass range
is highly desirable.
\end{quote}
\smallskip

\leftline{PACS codes: 12.39.Jh, 13.20.Jf, 13.25.Jx, 14.40.Rt}


\section{Introduction \label{sec:intro}}

The understanding of hadrons as bound states of colored quarks could
accommodate mesons as $q \bar q$ and baryons as $qqq$ states, but has remained
mute about the possible existence of more complicated color-singlet
combinations such as $qq \bar q \bar q$ (tetraquarks) or $q^4 \bar q$
(pentaquarks).  In the past dozen years or so, evidence has accumulated for
such combinations, but it has not been clear whether they are genuine bound
states with equal roles for all constituents, or loosely bound ``molecules''
of two mesons or a meson and a baryon, with quarks mainly belonging to one
hadron or the other.

A frequent agent for binding hadrons into molecules has been pion exchange
(\cite{Karliner:2015ina} and references therein) and in the case in which
nonstrange quarks are absent but strange quarks are present, possibly $\eta$
exchange \cite{Karliner:2016ith}.  A situation in which neither is possible
is a multi-quark state in which all the constituents are heavy ($c$ or $b$),
such as $c c \bar c \bar c$.  In comparison with states with two heavy and
two light quarks, a state such as $c b \bar c \bar b$ has a clear advantage
in binding, as the kinetic energy of its constituent quarks, scaling as
the inverse of their masses, is less.  The same is true for $cc \bar c \bar c$,
but not all configurations are allowed by the Pauli principle, so the situation
is less clear.  Starting more than 40 years ago \cite{Iwasaki:1975pv}, 
suggestions were made for producing and observing $cc \bar c \bar c$ states,
but there was no unanimity in whether these were above or below the lowest
threshold, $2 M(\eta_c) = 5967.2$ MeV, for a pair of $c \bar c$ mesons.  (See,
for example, Refs.\ \cite{Chao:1980dv,Ader:1981db,ZouZou:1986qh,%
Heller:1986bt,SilvestreBrac:1992mv,SilvestreBrac:1993ss,Semay:1994ht,%
Lloyd:2003yc,Chiu:2005ey,%
Barnea:2006sd,Berezhnoy:2011xy,Berezhnoy:2011xn,Heupel:2012ua,%
Wu:2016vtq,Chen:2016jxd,Bai:2016int}.)
We will present our own mass estimates, noting experimental strategies that
might be particularly appropriate for present-day and near-future searches.
We shall first discuss the lightest ``heavy tetraquarks," $cc \bar c \bar c$,
to be denoted generically as $X_{c c \bar c \bar c}$, as they are the
easiest to produce.  We will then present remarks on states $X_{b b \bar b \bar
b}$ containing $b$ quarks, which have a better chance of being narrow, and
will briefly mention mixed states $X_{bc \bar b \bar c}$.

Ingredients in estimating the mass of the lightest $X_{c c \bar c \bar c}$
state include the charmed quark mass, the color-electric force,
and the color-magnetic interaction leading to hyperfine splitting.  We discuss
the problems in evaluating each of them in Sec.\ II.  In contrast to previous
semi-empirical approaches (e.g.,
\cite{Karliner:2008sv,Karliner:2014opa,Karliner:2014gca} and references
therein), we utilize a relation between meson and baryon masses which allows
us to extrapolate to $QQ \bar Q \bar Q$ systems.  We discuss $c c \bar c \bar
c$ production in Sec.\ III and decay in Sec.\ IV.  Sec.\ V treats states
containing $b$ quarks.  Sec.\ VI contains remarks on tetraquarks with both $b$
and $c$ quarks, while Sec.\ VII summarizes.

\section{Estimating the ground-state $cc \bar c \bar c$ mass}

\subsection{Charmed quark mass}

In estimating the masses of baryons containing two heavy quarks
\cite{Karliner:2014gca}, we found the effective mass of the charmed quark in
mesons to be 1663.3 MeV, while in baryons it was found to be 1710.5 MeV.
The difference has been known for some time \cite{Gasiorowicz:1981jz}, and
is mirrored in a similar difference in constituent-quark masses in
mesons and baryons containing the light quarks $u$, $d$, and $s$.  It was
noted by Lipkin \cite{Lipkin:1978eh} that these effective masses differed
by approximately the same amount for strange and nonstrange quarks.  To see
this in a current context, we perform a least-squares fit to 5 ground-state
mesons and 8 ground-state baryons. The results, shown in Table \ref{tab:dif},
imply mass differences $m^b_q - m^m_q = 55.1$ MeV and
$m^b_s - m^m_s = 54.5$ MeV.  The square root of the average mean-squared
error in the fit is 6.72 MeV, for a six-parameter fit to thirteen data points. 

\begin{table}
\caption{Quark model description of ground-state hadrons containing $u,d,s$.
A least-squares fit to mesons gives $m^m_u = m^m_d \equiv m^m_q = 308.6$ MeV,
$m^m_s = 481.8$ MeV, $b/(m^m_q)^2 = 78.7$ MeV, while a fit to baryons gives
$m^b_u = m^b_d \equiv m^b_q = 363.7$ MeV, $m^b_s = 536.3$ MeV, and hyperfine
interaction term $a/(m^b_q)^2 = 49.3$ MeV.
\label{tab:dif}}
\begin{center}
\begin{tabular}{c c c c} \hline \hline
State (mass     & Spin &   Expression for mass   & Predicted  \\
in MeV)         &      & \cite{Gasiorowicz:1981jz} & mass (MeV) \\ \hline
$\pi(138)$      &  0   & $2 m^m_q - 6b/(m^m_q)^2$    & 145.3 \\
$\rho(775),\omega(782)$ & 1 & $2 m^m_q + 2b/(m^m_q)^2$ & 774.6 \\
$K(496)$        &  0   & $m^m_q + m^m_s - 6b/(m^m_q m^m_s)$ & 488.1 \\
$K^*(894)$      &  1   & $m^m_q + m^m_s + 2b/(m^m_q m^m_s)$ & 891.2 \\
$\phi(1019)$    &  1   & $2 m^m_s + 2b/(m^m_s)^2$ & 1028.1 \\ \hline
$N(939)$        & 1/2 & $3m^b_q - 3a/(m^b_q)^2$       &    943.3     \\
$\Delta(1232)$  & 3/2 & $3m^b_q + 3a/(m^b_q)^2$       &   1239.0    \\
$\Lambda(1116)$ & 1/2 & $2m^b_q + m^b_s - 3a/(m^b_q)^2$ &   1115.9  \\
$\Sigma(1193)$  & 1/2 & $2m^b_q + m^b_s+a/(m^b_q)^2-4a/m^b_q m^b_s$ & 1179.4 \\
$\Sigma(1385)$  & 3/2 & $2m^b_q + m^b_s+a/(m^b_q)^2+2a/m^b_q m^b_s$ & 1379.9 \\
$\Xi(1318)$     & 1/2 & $2m^b_s + m^b_q+a/(m^b_s)^2-4a/m^b_q m^b_s$ & 1325.4 \\
$\Xi(1530)$     & 3/2 & $2m^b_s + m^b_q+a/(m^b_s)^2+2a/m^b_q m^b_s$ & 1525.9 \\
$\Omega(1672)$  & 3/2 & $3m^b_s + 3a/(m^b_s)^2$       & 1677.0 \\ \hline \hline
\end{tabular}
\end{center}
\end{table}

\begin{table}
\caption{Quark model description of ground-state mesons and baryons containing
$u,d,s$, with universal quark masses for mesons and baryons but a constant
term $S= 165.1$ MeV added to baryon masses.  A least-squares fit gives
$m^m_u = m^m_d \equiv m_q = 308.5$ MeV, $m_s = 482.2$ MeV, $a/m_q^2 = 50.4$
MeV, $b/m_q^2 = 78.8$ MeV.
\label{tab:univ}}
\begin{center}
\begin{tabular}{c c c c} \hline \hline
State (mass     & Spin &   Expression for mass   & Predicted  \\
in MeV)         &      &                         & mass (MeV) \\ \hline
$\pi(138)$      &  0   & $2 m_q - 6b/(m_q)^2$    & 144.0 \\
$\rho(775),\omega(782)$ & 1 & $2 m_q + 2b/(m_q)^2$ & 774.8 \\
$K(496)$        &  0   & $m_q + m_s - 6b/(m_q m_s)$ & 488.0 \\
$K^*(894)$      &  1   & $m_q + m_s + 2b/(m_q m_s)$ & 891.6 \\
$\phi(1019)$    &  1   & $2 m_s + 2b/(m_s)^2$ & 1028.9 \\ \hline
$N(939)$        & 1/2 & $S + 3m_q - 3a/(m_q)^2$       &    939.4   \\
$\Delta(1232)$  & 3/2 & $S + 3m_q + 3a/(m_q)^2$       &   1242.1   \\
$\Lambda(1116)$ & 1/2 & $S + 2m_q + m_s - 3a/m_q^2$   &   1113.1   \\
$\Sigma(1193)$  & 1/2 & $S + 2m_q + m_s + a/m_q^2-4a/m_q m_s$ & 1185.7 \\
$\Sigma(1385)$  & 3/2 & $S + 2m_q + m_s + a/m_q^2+2a/m_q m_s$ & 1379.4 \\
$\Xi(1318)$     & 1/2 & $S + 2m_s + m_q + a/m_s^2-4a/m_q m_s$ & 1329.5 \\
$\Xi(1530)$     & 3/2 & $S + 2m_s + m_q + a/m_s^2+2a/m_q m_s$ & 1523.2 \\
$\Omega(1672)$  & 3/2 & $S + 3m_s + 3a/m_s^2$       & 1673.6 \\ \hline \hline
\end{tabular}
\end{center}
\end{table}

The near equality of nonstrange and strange quark mass differences between
mesons and baryons suggests a simpler fit with universal quark masses for
mesons and baryons but with a constant $S$ added to baryon masses.  The
results of this fit are shown in Table \ref{tab:univ}.  The quality of this
fit is nearly identical to that of the fit with separate quark masses for
mesons and baryons.  The square root of the average mean-squared error is 6.73
for a five-parameter fit to thirteen data points.

\begin{figure}
\includegraphics[width=0.95\textwidth]{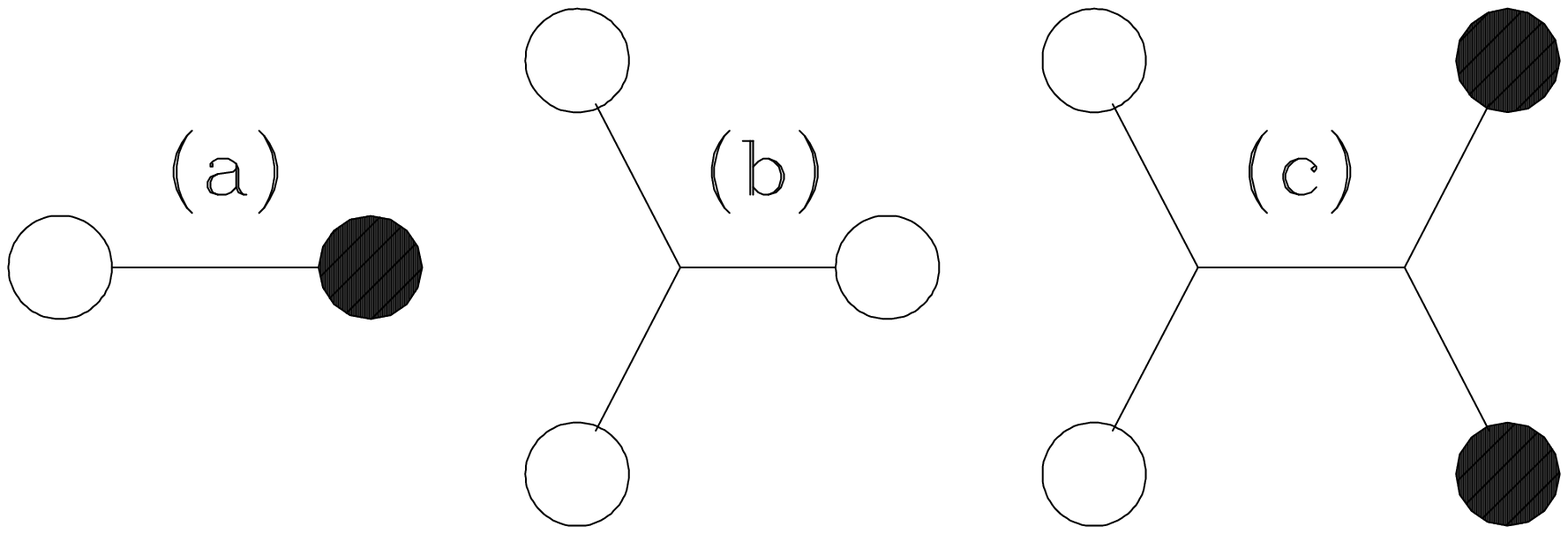}
\caption{QCD strings connecting quarks (open circles) and antiquarks
(filled circles).  (a) Quark-antiquark meson with one string and no
junctions; (b) Three-quark baryon with three strings and one junction;
(c) Baryonium (tetraquark) with five strings and two junctions.
\label{fig:jcts}}
\end{figure}

One can motivate the addition of a universal constant for baryon masses in a
QCD-string-junction picture \cite{Rossi:2016szw}.  A quark-antiquark meson
contains a single QCD string connecting a color triplet with an antitriplet.
A three-quark baryon contains three triplet strings, each leading to the
same junction.  Thus the added term $S$ may be thought of as representing
the contribution of two additional QCD strings and one junction.  (Fig.\
\ref{fig:jcts}.)

Now consider the baryonium (tetraquark) state consisting of two quarks and
two antiquarks, illustrated in Fig.\ \ref{fig:jcts}(c).  It contains five
QCD strings and two junctions, so one would expect an additional additive
contribution of $S$ with respect to a baryon or $2S$ with respect to a meson.
There will be additional contributions from binding effects and spin-dependent
interactions, like those considered in Ref.\ \cite{Karliner:2014gca}.

We estimate the charmed quark mass using $M(\Lambda_c) = S + 2 m_q + m_c
- 3a/m_q^2 = 2286.5$ MeV, obtaining $m_c = 1655.6$ MeV.  This is only slightly
different from the value obtained from mesons, and will be used henceforth.
As a cross-check, we calculate the mass of $\Sigma_c(2454)$ \cite{PDG} to be
\beq
M(\Sigma_c(2454)) = S + 2 m_q + m_c + \frac{a}{m_q^2} - \frac{4a}{m_q m_c}
 = 2450.5~{\rm MeV}~,
\eeq
to be compared with 2444 MeV in Ref.\ \cite{Karliner:2014gca}.

\subsection{Effects of interactions}

In this subsection we investigate the mass of a $cc \bar c \bar c$
state in which the $cc$ ($\bar c \bar c$) forms an S-wave color $3^*$ (3),
necessarily with spin 1 by the Pauli principle.  We follow a discussion
parallel to that in Ref.\ \cite{Karliner:2014gca}.  There we needed to
evaluate the mass of a $QQ$ color-$3^*$ spin-1 diquark.  We noted that
the binding energy for a $Q Q$ color-$3^*$, by QCD, was half that of a
$Q \bar Q$ color singlet.
(The picture of a diquark-antidiquark system $Q q \bar Q \bar q$ involving
two heavy quarks $Q$ and two light quarks $q$ has been used to describe
exotic states such as $X(3872)$, e.g., in Refs.\ \cite{Maiani:2004vq} and
\cite{Brodsky:2014xia}.)
The spin-averaged $1S$ charmonium mass, updating inputs based on \cite{PDG},
is \cite{Karliner:2014gca}
\beq
\bar M(c \bar c;1S) = [3M(J/\psi) + M(\eta_c)]/4 = 3068.5~{\rm MeV}~,
\eeq
so the $c \bar c$ (spin-averaged) binding energy in a color singlet is
\beq
B(c\bar c,1) = [3068.6 - 2(1655.6)]~{\rm MeV} = -242.7~{\rm MeV}~.
\eeq
and the $c c$ (spin-averaged) binding energy in a color $3^*$ is half that, or
$-121.3$ MeV.  The hyperfine interaction between two $c$ quarks in an S-wave
spin-1 color $3^*$ was estimated in Ref.\ \cite{Karliner:2014gca} to be
$a_{cc}/m_c^2 = 14.2$ MeV.  The effective mass of a $cc$ spin-1 color-$3^*$
diquark is then
\beq
M(cc,3^*) = [2(1655.6) -121.3 + 14.2]~{\rm MeV} = 3204.1~{\rm MeV}~.
\eeq

We next seek the binding energy of a $cc$ color-antitriplet diquark with a
$\bar c \bar c$ color-triplet antidiquark.  For this we will interpolate
between the 1S binding energies of $c \bar c$ and $b \bar b$, implicitly
assuming that the doubly-heavy diquarks are almost pointlike, as is the case
for $m_Q/\Lambda_{QCD} \to \infty$.  This approximation, while not perfect,
provides a concrete physically-motivated prescription for estimating the
strength of the binding between two diquarks.

We have already evaluated the 1S binding energy of $c \bar c$ to be $-242.6$
MeV.  To perform a comparable calculation for $b \bar b$ we retrace
steps in Ref.\ \cite{Karliner:2014gca}.  We first need an estimate for the
$b$ quark mass.  We use $M(\Lambda_b) = S + 2 m_q + m_b - 3a/m_q^2 = 5619.5$
MeV, obtaining $m_b = 4988.6$ MeV.  This is only slightly less than the value
of 5003.8 MeV obtained from mesons in Ref.\ \cite{Karliner:2014gca}.
It gives $M(\Sigma_b(5813))$ \cite{PDG} $ = S + 2 m_q + m_b + a/m_q^2 - 4a/(m_q
m_b) = 5808.6$ MeV, to be compared with 5805 MeV obtained in Ref.\
\cite{Karliner:2014gca}.  With 
\beq
\bar M(b \bar b;1S) = [3M(\Upsilon(1S))+M(\eta_b(1S))]/4=9445.0~{\rm MeV}~,
\eeq
the $b \bar b$ (spin-averaged) binding energy in a color singlet is
\beq
B(b \bar b,1) = [9445.0 - 2(4988.6)]~{\rm MeV} = -532.2~{\rm MeV}~.
\eeq

We can interpolate between this binding energy and that for charmonium to find
the binding energy $B$ for an antitriplet and triplet, each of mass 3204.1
MeV.  Assuming a power-law dependence of binding energy $B$ with constituent
mass $M$, $B_1/B_2 = (M_1/M_2)^p$, using $M_1 = 1655.6$ MeV, $M_2 = 4988.6$
MeV, $B_1 = -242.7$ MeV, $B_2 = -532.2$ MeV, we find $p = 0.7120$ and 
$B_3 = B_{(cc)(\bar c \bar c)} = -388.3$ MeV for $M_3 = M(cc,3^*) = 3204.1$ MeV.

Finally the spin-spin force between the spin-1 diquark $cc$ and the spin-1
antidiquark $\bar c \bar c$ may be estimated by interpolation between the
hyperfine splittings $\Delta M$ for $c \bar c$ and $b \bar b$ S-wave ground
states.  We assume $\Delta M_1/\Delta M_2 = (M_1/M_2)^q$, using $M_1 = 1655.6$
MeV, $M_2 = 4988.6$ MeV, $\Delta M_1 = 113.5$ MeV, $\Delta M_2 = 62.3$ MeV.
The power law is found to be $q = -0.5438$.  Then for a pair of spin-1/2 quarks
each of mass $M_3 = 3204.1$ MeV we would find $\Delta M_3 = 79.3$ MeV.

We assume the spin-dependent hyperfine splitting is of the form $A \langle
S_1 \cdot S_2 \rangle$, where
\beq
\langle S_1 \cdot S_2 \rangle = \frac{1}{2}[S(S+1) - S_1(S_1+1) - S_2(S_2+1)]~,
\eeq
where $S$ is the total spin and $S_{1,2}$ are the spins of the constituents.
For $S_1 = S_2 = 1/2$ the splitting between $S=1$ and $S=0$ states is $A$,
which we identify as the term $\Delta M_3 = 79.3$ MeV found above.  For
$S_1 = S_2 = 1$ the lowest-mass state, with $S=0$, lies $2A = 158.5$ MeV
below the value without hyperfine interaction.

Putting the terms together, we find the mass of the lowest-lying $c c \bar c
\bar c$ state in this configuration (with $J^{PC} = 0^{++}$) to be
$$
M(X_{cc \bar c \bar c}[0^{++}]) = 2S+2M_{cc} + B_{(cc)(\bar c \bar c)}
 + \Delta M_{HF}
$$
\beq
=  [2(165.1) + 2 (3204.1) - 388.3 - 158.5]~{\rm MeV} = 6191.5~{\rm MeV}~.
\eeq
This lies just below $J/\psi J/\psi$ threshold (6193.8 MeV), and cannot decay
to $J/\psi \eta_c$ (threshold 6080.5 MeV) by virtue of angular momentum and
parity conservation.  However, it can decay to $\eta_c \eta_c$ (threshold
5966.8 MeV), and thus is unlikely to be narrow.
We assign an error of $\pm 25$ MeV to this estimate, multiplying by two the
error \cite{Karliner:2014gca} expected in estimation of $QQq$ masses.

\subsection{Color-spin calculation}

The preceding analysis, based on string-junction physical picture,
suggests that the $c c \bar c \bar c$ tetraquark is likely to be
above $\eta_c \eta_c$ threshold. Since this is the crucial issue here,
it is useful to to do a cross-check with the help of another approach, 
namely color-spin $SU(6)$.

The dynamics of exotic combinations of quarks and antiquarks was examined
by combining the color $SU(3)$ and spin $SU(2)$ groups into a color-spin
$SU(6)$ \cite{Jaffe:1976ig,Jaffe:1976ih}.  [Particular attention was paid
to $qq \bar q \bar q$ baryon-antibaryon resonances \cite{Jaffe:1977cv},
as proposed in \cite{Rosner:1968si}.]  

Since the total chromoelectric interaction should not depend 
on the individual color groupings of the constituents
\cite{Lipkin:1981pq,Karliner:2006hf}, color-spin may be employed to
compare the binding energies of various $QQ \bar Q \bar Q$ states,
where $Q$ is a heavy quark which will be taken to be $c$ in the following.

Neglecting effects in which $QQ$ and $Q \bar Q$ have different relative wave
functions, the spin-dependent force $\Delta$ may be expressed in terms of
Pauli spin matrices $\vec{\sigma}$ and $SU(3)$ generators $\lambda^a$ ($a =
1, \ldots, 8$) as
$$
\Delta = - \sum_{a}^8 \sum_{i>j} \vec{\sigma}_i \cdot \vec{\sigma}_j \lambda_i^a
\lambda_j^a = 8N + \frac{1}{2}C_6({\rm tot}) -\frac{4}{3} S_{\rm tot}
(S_{\rm tot}+1)
$$
\beq
+ C_3(Q) + \frac{8}{3}S_Q(S_Q+1) -C_6(Q)
+ C_3(\bar Q) + \frac{8}{3}S_{\bar Q}(S_{\bar Q}+1) -C_6(\bar Q)~,
\eeq
where $N$ is the total number of quarks, and $C_3$ and $C_6$ are quadratic
Casimir operators of $SU(3)$ and $SU(6)$, whose relevant values are given
in Tables \ref{tab:cas3} and \ref{tab:cas6}.  (We use the normalization
of Ref.\ \cite{Jaffe:1976ih}.)

\begin{table}
\caption{Quadratic Casimir operators for $SU(3)$ representations.
\label{tab:cas3}}
\begin{center}
\begin{tabular}{c c} \hline \hline
$SU(3)$ rep    & $C_3$ \\ \hline
     1       &   0   \\
     3       &  16/3 \\
     6       &  40/3 \\
     8       &   12  \\ \hline \hline
\end{tabular}
\end{center}
\end{table}

\begin{table}
\caption{Quadratic Casimir operators for $SU(6)$ representations.
\label{tab:cas6}}
\begin{center}
\begin{tabular}{c c} \hline \hline
$SU(6)$ rep    & $C_6$ \\ \hline
     1       &   0   \\
     6       &  70/3 \\
     15      &  112/3 \\
     21      &  160/3 \\
     35      &   48   \\
    189      &   80   \\ \hline \hline
\end{tabular}
\end{center}
\end{table}

We first calculate $\Delta$ for the $\eta_c$, as we will be looking for
a configuration which is more deeply bound than two $\eta_c$s.  Here $N=2$,
while the deepest binding is achieved in an $SU(6)$ singlet with $C_6(1)=0$
and $S_{\rm tot} = 0$.  The terms describing individual quarks are
\beq
C_3(3) + \frac{8}{3}S_c(S_c+1) - C_6(c) = \frac{16}{3} + 2 - \frac{70}{3}
= -16
\eeq
with a similar term for $\bar c$, so
\beq
\Delta(\eta_c) = 8(2) - 2(16) = -16~;~~\Delta(2 \eta_c) = -32~.
\eeq

A corresponding calculation may be made in which $cc$ ($\bar c \bar c$) are
first combined into diquarks (antidiquarks).  The color-spin of $cc$ in
the ground state must be antisymmetric by Fermi statistics, so the
$SU(6)$ representation of the $cc$ ground state must be $15 = (6 \times 6)_A$.
The allowed $SU(6)$ representations are then $(15 \times \bar 15) = 1 + 35
+ 189$.  Here, as before, the deepest binding is achieved with $C_6({\rm tot})
= S_{\rm tot} = 0$, while the terms for individual quarks depend on which
$SU(3)$, $SU(2)$ representations of the $SU(6)$ 15-plet are chosen:
\beq
15 = (3^*,S=1) + (6,S=0)~.
\eeq
For $(3^*,S=1)$ we have
\beq
C_3(3^*) + \frac{8}{3}S(S+1) - C_6(15) = \frac{16}{3} + \frac{16}{3}
- \frac{112}{3} = -\frac{80}{3}
\eeq
with a similar term for antiquarks, while for $(6,S=0)$ we have
\beq
C_3(6) + \frac{8}{3}S(S+1) - C_6(15) = \frac{40}{3} - \frac{112}{3}
\eeq
which is less negative, and hence disfavored.\footnote{This confirms the
assumption used in Sec.\ B that the diquarks are anti-triplets of color and 
have spin 1.}

The final result for this configuration is
\beq
\Delta = 8(4) - 2 \frac{80}{3} = -\frac{64}{3}
\eeq
which is less deeply bound than two $\eta_c$s.  This supports our previous
estimate.  

As a caveat, one should note that the color-spin approach ignores the
distance between diquarks; everything depends only on the color-spin algebra.
From comparison of $\bar c c$ and $\bar b b$ quarkonia we know that this is an
oversimplification. In fact, the radii and the binding energies of these states
exhibit significant dependence on the quark mass, as utilized in Sec.\ B above.
So the color-spin approach should be viewed as qualitative, while the numbers
coming from the spin-junction approach are likely to be more reliable.

\subsection{Configuration mixing}

As noted above, the total chromoelectric interaction should not depend on the
individual color groupings of the constituents.
Thus, we may count ways of coupling two color triplets and two antitriplets
in several ways, but should end up with the same result.  In the previous
subsection we coupled $cc$ to a color $3^*$ and $\bar c \bar c$ to a color
$3$, then forming an overall singlet in the product $3^* \times 3 = 1 + 8$.
We could also have coupled $c c$ to a color $6$ and $\bar c \bar c$ to a color
$6^*$, then forming an overall singlet in the product $6 \times 6^* = 1 + 8
+ 27$.  An explicit calculation using Casimir operators verifies that the
chromoelectric interaction is the same for these two groupings.  Residual
interactions may split these two configurations.

A different grouping is obtained by combining each $c$ with a $\bar c$.  One
can form an overall singlet again in two ways.  Combining each $c$ with a $\bar
c$ in a color singlet, it is clear the final $(c \bar c) (c \bar c)$ state is a
singlet.  Two $\eta_c$s represent the lowest-lying $c c \bar c \bar c$ state in
this configuration.  Combining each $c$ with a $\bar c$ in a color octet, one
can form another overall singlet in the product $8 \times 8 = 1 + \ldots$.

\begin{figure}
\begin{center}
\includegraphics[width=0.8\textwidth]{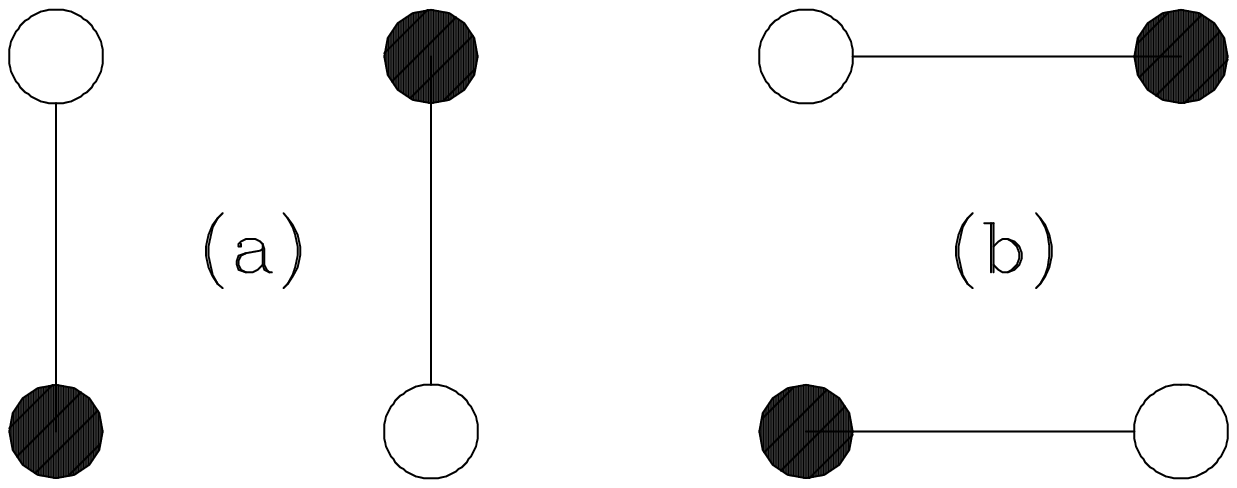}
\end{center}
\caption{Configurations of two quarks (open circles) and two antiquarks
(filled circles) at alternate vertices of a square.  QCD strings may run
either (a) vertically or (b) horizontally.
\label{fig:conf}}
\end{figure}

One can represent the two-fold nature of couplings to an overall singlet
by placing $c$ and $\bar c$ quarks at alternating vertices of a square, as
shown in Fig.\ \ref{fig:conf}.  One can draw QCD strings either (a) vertically
or (b) horizontally.  The incorporation of spins is simplest for the case in
which all spins are pointed in the same direction.  This represents two
parallel $J/\psi$ states coupled up to a total spin 2.  Tunneling between the
two configurations then will ensure mixing such that one eigenstate has a mass
greater than $2M(J/\psi)$ while the other has a mass less than $2M(J/\psi)$.
This is not enough to ensure a small decay width for the lighter state as
it may still be heavier than $2 M(\eta_c)$, but its $J^{PC} = 2^{++}$ will
force the $\eta_c \eta_c$ decay to be D-wave, and thus suppressed.  Its
decay to $\eta_c J/\psi$ will be forbidden by charge conjugation invariance.
One should bear in mind that the $2^{++}$ state might well not turn out to
be the lowest-mass $cc \bar c \bar c$ resonance; we have focused on it just
for the sake of simplicity.

If Fig.\ \ref{fig:conf}(a) represents a pair of $\eta_c$s, (b) will contain
admixtures of other states, so the effect of tunneling between the two
configurations is not as easy to evaluate.  Lattice gauge theory may be of
some help here.

A configuration related to that in Fig.\ \ref{fig:conf} is possible in
the binding of two positronium atoms to one another. It was first proposed by
Wheeler \cite{Wheeler:1946} and verified by a variational calculation in
Ref.\ \cite{Hylleraas:1947zza}.  Subsequent calculations (\cite{Kozlowski:1996}
and references therein) zeroed in on a binding energy of 0.435 eV, which is
$\sim 6\%$ of the binding energy of positronium [0.68 eV = (1/2)Ry].  Finally
this state was indeed produced by Cassidy and Mills \cite{Cassidy:2007};
Ref.\ \cite{Puchalski:2008jj} discusses its excitation and contains further
references.

The analogous situation for two quarkonium states is worth considering.
In the limit of very heavy quarks the binding is dominated by the
chromoelectric Coulomb force.  The existence of ``dipositronium'' thus
implies that an analog di-quarkonium state exists even though it need not have
the specific color network structure assumed for tetra-quarks.  Charmed quarks
are probably not heavy enough for this argument to hold, but we shall explore
it for bottom quarks in a subsequent section.

\section{Production of $c c \bar c \bar c$ states}

\subsection{States $X$ accompanying $J/\psi$ in $e^+ e^- \to J/\psi X$}

The strong production of a pair of heavy quarks $Q$ occurs at some cost,
depending on the process.  As an example consider the reaction $e^+ e^-
\to c \bar c$, whose cross section far above threshold is 4/3 that for
muon pair production:  at a center-of-mass energy $\sqrt{s}$,
\beq
\sigma(e^+ e^- \to c \bar c) = \frac{4}{3} \frac{4 \pi\alpha^2}{3 s}\left(
1 - \frac{4 m_c^2}{s} \right)^{1/2} \left(1 + \frac{2 m_c^2}{s} \right)
\simeq 1~{\rm nb~at}~\sqrt{s} = 10.6~{\rm GeV}~.
\eeq
In $e^+ e^- \to J/\psi X$, the mass $M(X)$ shows peaks at states with $J^{PC} =
0^{\pm +}$:  notably $\eta_c(2984)$, $\chi_{c0}(3415)$, $\eta_c(3639)$, and
$X(3940)$ \cite{Abe:2001za,Abe:2002rb,Abe:2007jna}, as well as a continuum
above $D\bar D$ threshold.  The inclusive cross section for $e^+ e^- \to J/\psi
c \bar c$ at $\sqrt{s}=10.6$ GeV is about 0.9 pb \cite{Abe:2001za,Abe:2002rb},
and dominates the inclusive $J/\psi$ production cross section
\cite{Abe:2002rb}:
\beq
\frac{\sigma(e^+ e^- \to J/\psi c \bar c)}{\sigma(e^+ e^- \to J/\psi X)} =
0.59^{+0.15}_{-0.13}\pm 0.12
\label{eq:inclusiveJpsi}
\eeq
Thus, very roughly, the probability for producing a $c \bar c$ pair
when one is already present is about $10^{-3}$.  We may use this figure in
comparing, say, production of the tetraquark $c \bar c c \bar c$ with that
of a typical quarkonium state.

The somewhat counterintuitively large ratio on the right-hand side of 
Eq.~(\ref{eq:inclusiveJpsi}) can be understood as follows.  If $J/\psi$ is 
produced at high $e^+ e^-$ energy, its $c$ and $\bar c$ are unlikely to have
come from the same primary photon, so there tends to be another $c \bar c$ pair
around.  A smaller probability is associated with a final $c$ and $\bar c$
both connected to the initial photon, with light hadrons coupling to the $c$
and/or $\bar c$ by gluons.

\subsection{Inclusive double charm production at the LHC}

The LHCb Collaboration has measured prompt charm production at the CERN Large
Hadron Collider (LHC) \cite{Aaij:2016jht}.  In the kinematic range studied,
$2.0 < y < 4.5$ and $1 < p_T < 8$ GeV/$c$, they report a total cross section
for charm production of about 1 mb at $\sqrt{s} = 5$ TeV (see their Fig.\ 10)
and about twice that at $\sqrt{s} = 13$ TeV.  Doubling these values for the
contribution of $-4.5 < y < -2.0$ and accounting for central production
contributions of similar order, one estimates $\sigma(pp \to c \bar c
X) \simeq 5$--10 mb at $\sqrt{s} = 13$ TeV.  Now one applies the estimate of
the previous subsection, that an additional charm pair appears with a
probability of about $10^{-3}$, to estimate
\beq
\sigma(pp \to c c \bar c \bar c) \simeq 5\!-\!10~\mu{\rm b}~{\rm at}~
\sqrt{s} = 13~{\rm TeV}~.
\eeq

These four quarks would form a tetraquark state with low probability.  If
produced by an initial gluon, each $c \bar c$ pair has a low effective mass,
typically not more than several (e.g., 4) times $m_c$, whereas if correlated
in a tetraquark state the relative effective mass of each pair should be within
$\Lambda_{\rm QCD} \sim 200$ MeV of $2m_c$.  Thus we estimate a suppression
factor of $(\sim 6/0.2)^2 \sim 10^3$ from demanding these correlations.  In
addition, the two $c \bar c$ pairs must be close to one another in rapidity
space to be accommodated in a resonant state.  Let us assume this costs another
suppression factor of at least ten.  Then we would obtain a cross section for
tetraquark production of no more than 1 nb at $\sqrt{s} = 13$ TeV, and less
at lower energies.  One might expect this estimate to be accurate give or take
a factor of three.

This crude estimate can be checked by noting the cross section for double
charmonium production, which has been measured by ATLAS \cite{Aaboud:2012fzt},
CMS \cite{Khachatryan:2014iia}, and LHCb \cite{Aaij:2011yc}.  These results are
summarized in Table \ref{tab:doublepsi}.  Very roughly, one may quadruple the
LHCb result to account for the full rapidity range $-4.5 < y < 4.5$ to estimate
\beq
\sigma(pp \to J/\psi J/\psi X) \simeq 20~{\rm nb~at}~\sqrt{s} = 7~{\rm TeV}~.
\eeq
Now one may use an estimate (see Sec.\ III B of Ref.\ \cite{Berezhnoy:2011xn})
that the ratio of tetraquark to $J/\psi$ pair production by two gluons is 3.5\%
to conclude that the tetraquark production cross section in proton-proton
collisions at $\sqrt{s} = 7$ TeV is about 0.7 nb.  This is consistent
with our very rough estimate above.


\begin{table}
\caption{Double $J/\psi$ production at the LHC.
\label{tab:doublepsi}}
\begin{center}
\begin{tabular}{c c c c c} \hline \hline
Experiment & $\sqrt{s}$ & $y$ range & $p_T$ range & $\sigma$ \\ \hline
ATLAS \cite{Aaboud:2012fzt} & 8 TeV & $|y|<2.1$ & $>8.5$ GeV/$c$
 & 160$\pm$12$\pm$14$\pm$2$\pm$3 pb \\
CMS \cite{Khachatryan:2014iia} & 7 TeV & $|y|<1.2$ & $>6.5$ GeV/$c$ & \\
                               & 7 TeV & $1.2<|y|<1.43$ & (a) & \\
       & 7 TeV & $1.43<|y|$ & $>4.5$ GeV/$c$ & 1.49$\pm$0.07$\pm$0.13 nb \\
LHCb \cite{Aaij:2011yc} & 7 TeV & $2.0<y<4.5$& $<10$ GeV/$c$
 & 5.1$\pm$1.0$\pm$1.1 nb\\
\hline \hline
\end{tabular}
\end{center}
\centerline{(a) $p_T$ scaled linearly from 6.5 to 4.5 GeV/$c$}
\end{table}

\section{Decays of $c c \bar c \bar c$ states}

A number of final states are accessible to decays of a $c c \bar c \bar c$
resonance.  Some of these are summarized in Table \ref{tab:cfs}.  Here $\ell$
stands for any charged lepton $(e,\mu,\tau)$, $h$ stands for any hadron, and
$g$ stands for a gluon.  Invariances under spin, parity, and charge conjugation
may suppress certain final states.

\begin{table}
\caption{Some final states accessible to decays of a $c c \bar c \bar c$
resonance $X_{c c \bar c \bar c}$.
\label{tab:cfs}}
\begin{center}
\begin{tabular}{|c|c|c|} \hline \hline
Subprocess & Resulting & Maximum kinetic\\
& final state & energy available \\ \hline
$2(c\bar c \to \gamma \gamma)$ & $\gamma \gamma$ &  $M(X_{c c \bar c \bar
c})$  \\ 
\hline
$c_1 \bar c_1 \to \gamma (\gamma)$, & & \\
$c_2 \bar c_2 \to \ell^+ \ell^-$ & $\gamma (\gamma) \ell^+ \ell^-$
 & $M(X_{c c \bar c \bar c}) - 2M(\ell)$ \\ 
\hline
$c_1 \bar c_1 \to \ell_1^+ \ell_1^-$, & & \\
$c_2 \bar c_2 \to \ell_2^+ \ell_2^-$ & $\ell_1^+\ell_1^-\ell_2^+\ell_2^-$ &
 $M(X_{c c \bar c \bar c}) - 2M(\ell_1) - 2M(\ell_2)$ \\ 
\hline
$c_1 \bar c_1 \to \gamma (\gamma)$, & & \\
$c_2 \bar c_2 \to h^+ h^-$ & $\gamma (\gamma) h^+ h^-$ 
& $M(X_{c c \bar c \bar c}) -2M(h)$ \\
\hline
$2(c \bar c) \to gg$ & Light hadrons & $M(X_{c c \bar c \bar c}) -2M(\pi)$ \\
\hline
$c \bar c \to \gamma$ & $\eta_c \gamma$ \,or\, $J/\psi \gamma$ 
& $M(X_{c c \bar c \bar c}){-}M(\eta_c)$ \,or\,
$M(X_{c c \bar c \bar c}){-}M(J/\psi)$ \\
\hline
$c \bar c \to \gamma$ & $D \bar D \gamma$ & $M(X_{c c \bar c \bar c}) -2M(D)$ \\
\hline
$c \bar c \to \ell^+\ell^-$ & $\ell^+\ell^- D \bar D$ 
& $M(X_{c c \bar c \bar c}) - 2M(\ell)-2M(D)$ \\
\hline
$c \bar c \to q \bar q$ & $D \bar D +\hbox{anything}$
& $M(X_{c c \bar c \bar c}) -2M(D)$ \\
\hline
Rearrangement & $2 \eta_c$ 
& $M(X_{c c \bar c \bar c}) - 2M(\eta_c)$ \\ \hline \hline
\end{tabular}
\end{center}
\end{table}

If the lowest $c \bar c c \bar c$ tetraquark mass exceeds $2M(\eta_c) =
(5967.2 \pm 1.4)$ MeV, such a state will decay primarily into the decay
products of any open charmonium pair channel.  Thus, for example, a 6000 MeV
$c \bar c c \bar c$ tetraquark with $J^{PC} = 0^{++}$
may be expected to have primarily the decay products of two $\eta_c$ mesons.

\begin{figure}
\begin{center}
\includegraphics[width=0.7\textwidth]{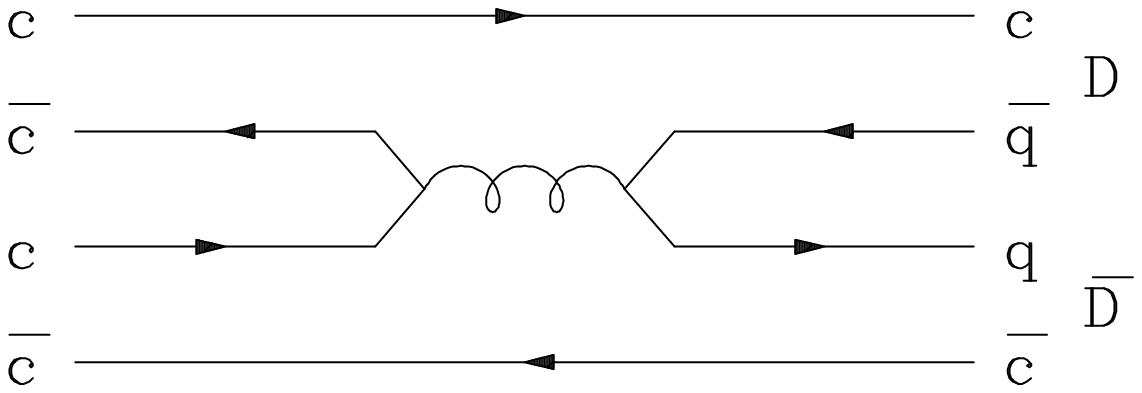}
\end{center}
\caption{Lowest order process governing decay of a $c c \bar c \bar c$
resonance whose mass is below $2M(\eta_c)$.
\label{fig:oneg}}
\end{figure}

If $M(X_{c c \bar c \bar c}
[0^{++}])$ is less than $2M(\eta_c)$, the main decay products will involve the
subprocess $c \bar c \to g^* \to q\bar q$, illustrated in Fig.\ \ref{fig:oneg}.
All other processes are higher-order in the strong interactions or involve at
least one electromagnetic interaction.

The rate for the process illustrated in Fig.\ \ref{fig:oneg} may be crudely
estimated by comparing it with the rate for leptonic decay of the $J/\psi$
\cite{PDG}:
\beq \label{eqn:clept}
\Gamma(J/\psi \to e^+ e^-) = (5.55 \pm 0.24 \pm 0.02)~{\rm keV}~.
\eeq
Leaving aside group-theoretic factors of order 1, and assuming the wave
function at the origin in the tetraquark for $c \bar c \to g^* \to q \bar q$
is about the same as for $c \bar c \to \gamma^* \to e^+ e^-$, the rate for the
process of Fig.\ \ref{fig:oneg} is approximately $(\alpha_s/\alpha)^2$ times
that of the leptonic decay process (\ref{eqn:clept}).  Taking $\alpha_s = 0.35$
at a scale $m_c$ (cf.\ Ref.\ \cite{Ackerstaff:1998yj} for a measurement at
$m_\tau$), we have $(\alpha_s/\alpha)^2 \simeq 2300$ or
\beq \label{eqn:gtotc}
\Gamma(X_{c c \bar c \bar c})~ \simeq 13~{\rm MeV}~.
\eeq

Other less significant modes include
\beq
c \bar c c \bar c \to g g \to ~{\rm light~hadrons}~{\rm or}~
c \bar c c \bar c \to \gamma^* \gamma^*~,
\eeq
where the virtual photons will materialize into lepton or hadron pairs.  The
relative branching fractions to gluon or virtual photon pairs will depend on
details of color-spin groupings:  A color-octet $c \bar c$ pair with $J=1$
will decay to a gluon, while a color-singlet $c \bar c$ pair with $J=1$ will
decay to a virtual photon.  Decays of $J=0$ $c \bar c$ pairs will involve
more than one gluon and/or virtual photon.

Although the expected branching fractions are likely to be small, the channels
in which both virtual photons materialize as lepton pairs are worth
investigating.  One should see final states of $2\tau^+2\tau^-,\tau^+\tau^-
\mu^+\mu^-,\tau^+\tau^-e^+e^-,2\mu^+2\mu^-,\mu^+\mu^-e^+e^-$, and $2e^+2e^-$ in
the ratio 1:2:2:1:2:1.  In that case there will also be channels
in which one or both virtual photons materialize as hadrons containing
$u, d,$ and $s$ quarks, with well-defined branching ratios.

A crude estimate of the 4-lepton branching fraction of a $c_1 c_2 \bar c_1 \bar
c_2$ tetraquark may be made as follows. The partial width for the $c_1\bar c_1$
pair in a color singlet $^3S_1$ ground state to decay to an $e^+e^-$ pair is
just $\Gamma(J/\psi \to e^+ e^-) = 5.5$ keV [Eq.\ (\ref{eqn:clept})]. In a
tetraquark with $J^{PC} = 0^{++}$ (expected to be lightest) this leaves the
remaining $c_2 \bar c_2$ pair also in a color singlet $^3S_1$ state.  If $c_1
\bar c_1$ is sufficiently off-shell, there will remain enough phase space for
$c_2 \bar c_2$ to decay not only to a lepton pair [with partial width
(\ref{eqn:clept})] but also a pair of charmed mesons [if the effective mass of
$c_2 \bar c_2$ is above $2M(D) = 3.73$ GeV].  Comparing these two channels, we
see that the charmed meson pair decay is likely to have a partial width of
order tens of MeV, or about $10^3$ that of the decay to a pair of charged
leptons, unless the penalty for $c_1 \bar c_1$ being off-shell is very great.
In that case the factor of $10^3$ might be replaced by a quantity as small as
unity.  Taking account of the total width estimate (\ref{eqn:gtotc}), one then
estimates
\beq
{\cal B}(X_{c c\bar c\bar c}[0^{++}] \to \ell_1^+ \ell_1^- \ell_2^+
\ell_2^-) = (1~{\rm to}~10^{-3}) (11~{\rm keV})/(13~{\rm MeV}) = {\cal O}
(10^{-3}~{\rm to}~10^{-6})~.
\eeq
The higher branching ratio would very likely involve at least one lepton pair
with mass $J/\psi$.  If there is a $c c \bar c \bar c$ tetraquark below
$2M(\eta_c)$ (unlikely in our opinion), the cross section for its production
and observation in the four-lepton mode at the LHC (13 TeV) is estimated
to lie in the range of 1 fb -- 1 pb.

\section{States containing bottom quarks}

\subsection{Ground state mass estimate}

The threshold for a ``fall-apart'' decay of a $b b \bar b \bar b$ tetraquark
with $J^{PC} = 0^{++}$ or $2^{++}$ is $2M(\eta_b) = (18{,}798 \pm 4.6)$ MeV.
For one estimate of the mass of the lowest $b b \bar b \bar b$ state, we can
repeat the $cc \bar c \bar c$ calculation which envisioned $cc$ diquarks
interacting with $\bar c \bar c$ antidiquarks.  We already estimated
$\bar M(b \bar b;1S) = 9445.0$ MeV and $B(b \bar b;1) = -532.2$ MeV, so
$B(bb,3^*) = -266.1$ MeV.  Taking into account a small hyperfine contribution
\cite{Karliner:2014gca} of $a_{bb}/m_b^2 = 7.8$ MeV, this implies $M(bb,3^*) =
9718.9$ MeV.  Using the power-law relation $B_3/B_2 = (M_3/M_2)^{0.712}$
employed previously, for $B_2 = - 532.2$ MeV, $M_2 = 4988.6$ MeV, and $M_3 =
9718.9$ MeV, we obtain $B_3 = -855.7$ MeV for the binding energy between the
$bb$ diquark and the $\bar b \bar b$ antidiquark.

The evaluation of the hyperfine interaction is similarly straightforward.
Using $\Delta M_3/\Delta M_2 = (M_3/M_2)^{-0.5438}$ and $\Delta M_2 = 62.3$
MeV, we find $A = 43.35$ MeV and $-2A = -86.7$ MeV.  The final calculation
gives
\bea
M(X_{bb\bar b \bar b}[0^{++}]) & = & 2S + 2 M(bb,3^*) + B_{(bb)(\bar b \bar b)}
 + \Delta M_{HF} \nonumber \\ 
& = & [2(165.1) + 2(9718.9) -855.7 - 86.7]~{\rm MeV} = 18{,}825.6~{\rm MeV}~.
\eea
As in Sec.\ II B, we assign an error of $\pm 25$ MeV to this estimate,
corresponding to twice the error assigned in Ref.\ \cite{Karliner:2014gca}
to estimates of $QQq$ masses.
This lies 95.0 MeV below $2M(\Upsilon(1S))$, 33.7 MeV below $M(\Upsilon(1S)) +
M(\eta_b)$, and 27.6 MeV above $2M(\eta_b)$.  This is to be compared with the
estimate in Sec.\ II B of $M(X_{cc\bar c \bar c}[0^{++}]) = 6191.5$ MeV, 224.3
MeV above $2M(\eta_c) = 5967.2$ MeV.  Thus there is a chance that the lowest
$bb \bar b \bar b$ state is narrow enough to be visible in a mode other than
those coming from the decays of individual $\eta_b$ components.

The example of dipositronium discussed in the previous section can be applied
to the case of the bottom quark. With $m_b\approx 5$ GeV and an effective value
of $\alpha_s = 0.35$ for the QCD Coulombic interaction, the binding energy
for two spin-triplet states (neglecting Casimir operators of order unity) will
be
\beq
B[\Upsilon(1S) \Upsilon(1S)] = (1/4) \, m_b \,\alpha_s^2 \cdot 0.06 \simeq
9~{\rm MeV}~.
\eeq
This state is above $2M(\eta_b)$, so that will be its dominant decay, but the
discussion confirms the earlier estimate that two $\Upsilon(1S)$ can form a
molecule, even if only weakly bound.

\subsection{Production of the lowest $b b \bar b \bar b$ state}

Recently the CMS Collaboration \cite{Khachatryan:2016ydm} has observed $38\pm7$
events of $\Upsilon(1S)$ pairs produced with an integrated luminosity of 20.7
fb$^{-1}$ at $\sqrt{s} = 8$ TeV, each decaying to $\mu$ pairs.  The reported
fiducial cross section, with each $\Upsilon(1S)$ required to have rapidity $|y|
<2$, is $68.8\pm12.7~({\rm stat}) \pm 7.4~({\rm syst}) \pm 2.8~({\cal B})$ pb.
It is estimated in one theoretical calculation \cite{Berezhnoy:2012tu} that
about 30\% of this value is due to double-parton scattering and another 12 pb
is due to feed-down from $\Upsilon(2S) \Upsilon(1S)$ production, leaving
36 pb for $\Upsilon(1S)$ pair production without feed-down.  In analogy with
our discussion of the relation between $J/\psi$ pair and $c c \bar c \bar c$
tetraquark production, we expect the latter to be a few percent of the
former, implying (at 8 TeV)
\beq \label{eqn:sigbtq}
\sigma(pp \to X_{b b \bar b \bar b}) \simeq 1~{\rm pb}~,
\eeq
or about twice that at 13 TeV.  (The LHCb Collaboration \cite{Aaij:2016avz} has
found that the rate for single-$b$ production roughly doubles from 7 to 13
TeV.)

\subsection{Decays of the lowest $b b \bar b \bar b$ state}

The predicted mass of the lowest $b b \bar b \bar b$ state is only about
$28 \pm 25$ MeV above $2 M(\eta_b)$.  This suggests that a hadronic decay
into two $\eta_b$ mesons, followed by their individual decays, may not be the
only decay mode of $X_{b b \bar b \bar b}$.  If its mass is actually
below $2 M(\eta_b)$, decay occurs when each $b$ annihilates a $\bar b$,
or when one $b$ annihilates a $\bar b$ and the other $b \bar b$ pair emerges
as a pair of $B$-flavored mesons in the manner akin to Fig.\
\ref{fig:oneg}.  In analogy with our calculation for charm, we can compare
the expected rate for this process with the leptonic width \cite{PDG}
\beq \label{eqn:blept}
\Gamma(\Upsilon(1S) \to e^+ e^-) = (1.340 \pm 0.018)~{\rm keV}.
\eeq
Taking $\alpha_s(m_b) = 0.22$ \cite{PDG}, we have $(\alpha_s/\alpha)^2 \simeq
900$ or
\beq \label{eqn:gtotb}
\Gamma(X_{b b \bar b \bar b})~ \simeq 1.2~{\rm MeV}~.
\eeq
Assuming that the lowest $b_1 b_2 \bar b_1 \bar b_2$ tetraquark decays with
$b_1 \bar b_1 \to \ell_1^+ \ell_1^-$ with partial width approximately equal to
(\ref{eqn:blept}), and partial width for $b_2 \bar b_2$ decay ranging from
(\ref{eqn:blept}) to tens of MeV, one predicts
\beq
{\cal B}(X_{b b \bar b \bar b}[0^{++}] \to \ell_1^+ \ell_1^- \ell_2^+
\ell_2^-) = (1~{\rm to}~10^{-4}) (2.7~{\rm keV})/(1.2~{\rm MeV}) = {\cal O}
(2 \times 10^{-3}~{\rm to}~2 \times 10^{-7})~.
\eeq
This implies a cross section for the four-lepton observation of a $b b \bar b
\bar b$ tetraquark:
\beq \label{eqn:sigbbtq}
\sigma(pp \to X_{b b \bar b \bar b}[0^{++}] \to \ell_1^+ \ell_1^-
\ell_2^+ \ell_2^-) \le 4~{\rm fb~~(LHC, 13 TeV)}~,
\eeq
where the upper limit is attained only if there is not significant competition
from the decay mode
\beq
X_{b b \bar b \bar b}[0^{++}] \to \ell^+ \ell^-~B \bar B X~.
\eeq
At 7 or 8 TeV one would expect about half this, or 2 fb.

\section{Remarks on mixed states}

If heavy quarks in a tetraquark are produced in quark-antiquark pairs, one
might expect tetraquarks of the form $b \bar b c \bar c$ to be much more
abundant than $bb \bar c \bar c$ or $c c \bar b \bar b$ tetraquarks.  The
following remarks thus apply only to $b \bar b c \bar c$ states.  One would
expect their production cross section in a hadronic reaction to be
intermediate between that of $c c \bar c \bar c$ and $bb \bar b \bar b$.
One would probably not expect the lowest-mass $b \bar b c \bar c$ state to
lie below $M(\eta_c)+M(\eta_b)$.  In that unlikely case, however, there are
fewer opportunities for heavy-quark annihilation than for $c c \bar c \bar c$
or $b b \bar b \bar b$, as each quark has only one antiquark with which to
annihilate.

The dominant decay will then be annihilation of a
single heavy quark pair into a light quark pair (analogous to the process in
Fig.\ \ref{fig:oneg}), leading to a total width of several MeV.  Production
cross sections at the LHC would be several tens of pb.  Expected branching
fractions to a four-lepton final state could be as large as $10^{-3}$ but
could be several orders of magnitude smaller if the decays
\beq
X_{b c \bar b \bar c}[0^{++}] \to \ell^+ \ell^-~D \bar D X~,~~
 \ell^+ \ell^-~B \bar B X
\eeq
played a dominant role.

\section{Conclusions}

We have estimated the mass of the lowest-lying $c \bar c c \bar c$ tetraquark
and find it unlikely to be less than twice the mass of the lowest charmonium
state $\eta_c$.  In that unlikely case, however, the decay may proceed by
annihilation of each $c \bar c$ pair as long as each is in a $J=1$ state.
In that case, one expects final states of hadrons from pairs of intermediate
gluons, and of hadrons or leptons from pairs of intermediate virtual photons.
Similar arguments apply to the heavier tetraquarks $b \bar b c \bar c$ and
$b \bar b b \bar b$.
The predicted masses of the lowest-lying states are
$M(X_{cc\bar c \bar c}[0^{++}]) = 6{,}192 \pm 25$ MeV
and
$M(X_{bb\bar b \bar b}[0^{++}]) = 18{,}826 \pm 25$ MeV,
for the charmed and bottom tetraquarks, respectively.
The proximity of the predicted $(bb)(\bar b \bar b)$ mass to $2M(\eta_b)$
suggests that if we have overestimated it by an amount comparable to our
uncertainty, its decays to a pair of real or virtual photons or a pair of
gluons may stand a chance of being observable.  Other estimates of resonant
$c c \bar c \bar c$ and $b b \bar b \bar b$ masses, summarized in Table
\ref{tab:mass_pred}, give mixed signals as to whether the lightest state is
above or below the mass of the lightest quarkonium pair.

Searches in the four-lepton and $\ell^+ \ell^- B \bar B$ final states have
been performed at the LHC \cite{Aad:2015oqa,Khachatryan:2017mnf}.  These are
devoted to the search for the standard-model Higgs boson decaying into two
light pseudoscalars $a$, which then decay to such final states as $\mu^+
\mu^-,~\tau^+\tau^-,$ and $b \bar b$.  These are ideal samples for the searches
advocated here.

\begin{table}[t]
\def\upstrut{\vrule height 2.5ex depth 0.0ex width 0pt}
\caption{Predictions for the mass of the $Q Q \bar Q \bar Q$ tetraquark
\label{tab:mass_pred}}
\begin{center}
\begin{tabular}{c c c} \hline \hline
\upstrut
Reference & $M(X_{c c \bar c \bar c})$ & $M(X_{b b \bar b \bar b})$ \\
          &            (MeV)           &        (MeV)           \\ \hline
This work & $6{,}192\pm25~(0^{++})$    & $18{,}826\pm 25~(0^{++})$             \\
\cite{Iwasaki:1975pv} & $\sim$6,200 & -- \\
\cite{Lloyd:2003yc} & 6,908 & -- \\
\cite{Barnea:2006sd} & 6,038 & -- \\
\cite{Berezhnoy:2011xn} & 5,966$(0^{++})$ & 18,754$(0^{++})$ \\
\cite{Berezhnoy:2011xn} & 6,051$(1^{+-})$ & 18,808$(1^{+-})$ \\
\cite{Berezhnoy:2011xn} & 6,223$(2^{++})$ & 18,916$(2^{++})$ \\
\cite{Heupel:2012ua} & $5{,}300\pm 500$ & -- \\
\cite{Wu:2016vtq} & 5,617--6,254 & 18,462--18,955\\
\cite{Chen:2016jxd} & $6{,}440\pm150$ & $18{,}450\pm150$\\
\cite{Bai:2016int}$^*$\kern-0.5em & -- & $18{,}690\pm30$ \\ \hline \hline
\end{tabular}
\end{center}
\vskip -0.3cm
\centerline{\footnotesize $^*$Appeared after the first version of the
current work}
\vskip -0.7cm\strut
\end{table}

\section*{Acknowledgements}

We thank Sheldon Stone for stimulating us to address this question and for
helpful comments.  We are also grateful to Matt Strassler and Carlos
Wagner for encouraging remarks, and to Xiang Liu and Alexey Luchinsky for
informing us of earlier work.  J.L.R.
thanks Tel Aviv University for hospitality during the inception of this work,
which was supported by the U.S. Department of Energy, Division of High Energy
Physics, Grant No.\ DE-FG02-13ER41958.

\end{document}